\documentclass[notoc]{JHEP}
\newcommand{\nn}{\noindent}
\newcommand{\no}{\nonumber\\}
\newcommand{\be}{\begin{equation}}
\newcommand{\ee}{\end{equation}}
\newcommand{\ba}{\begin{eqnarray}}
\newcommand{\ea}{\end{eqnarray}}
\newcommand{\ci}[1]{\cite{#1}}
\newcommand{\bi}[1]{\bibitem{#1}}
\newcommand{\la}[1]{\label{#1}}
\def\gl#1{(\ref{#1})}

\title{Composite Two-Higgs Models  and Chiral 
Symmetry Restoration}

\author{A. A. Andrianov\thanks{Supported by Grants DFG  436 RUS 113/227/5,
 RFBR 98-02-18137, GRACENAS  6-19-97 and by Generalitat de Catalunya, 
Grant PIV 1999.},\\  Departament 
d'Estructura i Constituents de la Mat\`eria,\\
 Universitat de Barcelona,\\
Diagonal, 647, 08028 Barcelona, Spain,}
\author{V. A. Andrianov\thanks{Supported by Grants DFG  436 RUS 113/227/5,
 RFBR 98-02-18137, GRACENAS  97-40-1.6-7},\\
 Department of Theoretical Physics,\\ 
                 Institute of Physics, St.-Petersburg State University,\\
                                198904 St.-Petersburg, Russia,}
\author{R. Rodenberg\thanks{Supported by DFG Grant  436 RUS 113/227/5.},
\\ Abteilung f\"ur Theoretische Elementarteilchen Physik,\\
        III. Physikalisches Institut,    Physikzentrum, RWTH-Aachen,\\
                                            D-52056, Aachen, Germany}

\abstract{
The effective quark models  with 
quasilocal interaction are used for description of two composite Higgs doublets,   in strong coupling 
(tricritical) regime below the compositeness scale $\Lambda_{C}$.
The low energy effective action of Two-Higgs Doublet Standard Model (2HD SM) is obtained in
the large $N_c$ and large-log approximation.
The two-point correlators of scalar and pseudoscalar Higgs fields are derived for investigation of how 
the chiral symmetry is broken.
The comparison of their asymptotics  at high energies allows to realize
the chiral symmetry restoration characteristic for the QCD-like  models
and thereby to make hints on the existence of new physical phenomena in the
TeV energy region.} 
\preprint{PITHA 99/6\\26.February 1999}
\dedicated{ In Memory of Alexei A. Anselm}
\begin{document}
\section{Introduction: definition of 2HQQM}
\hspace*{3ex} This paper we dedicate to the memory of outstanding
Russian physicist Alexei A. Anselm whose teaching and influence during
many years was indispensable in our understanding of the Higgs
phenomenon \ci{ans}.

One of the minimal extensions of the Standard Model (SM) is the Two
Higgs Doublet Model (2HDM) \ci{lee,gun} which has two complex doublets
of Higgs bosons instead of only one in SM.  The general 2HDM allows
too strong flavour-changing neutral currents (FCNC) \ci{gw} as
compared to the phenomenology of electroweak decays \ci{at} .  One
possibility to suppress FCNC is to couple the fermions only to a fixed
combination of the two Higgs doublets and its charge conjugated one
which is know as a Model I \ci{gun}. There is another possibility to
couple the first Higgs doublet to down-type quarks while the second
one to up-type quarks, which is known as a Model II \ci{gw} .  The
physical content of the Higgs sector includes a pair of CP-even
neutral scalar Higgs bosons, $H^0$ and $h^0$, a CP-odd neutral
pseudoscalar Higgs boson $A$ and a pair of charged Higgs bosons
$H^\pm$. The mass spectrum of Models I and II has been extensively
studied theoretically \ci{lee,gun,krod,sher} and bounded from the
phenomenology of EW interactions at available high energies (for
recent analyses, see \ci{lang,degr} and refs. therein).

If Higgs bosons are composite  and their masses are created by a
mechanism \ci{njl} of spontaneous chiral symmetry breaking 
\ci{bard,marc,bar,ll}
below a scale of compositeness $\Lambda$  one can require the chiral
symmetry restoration (CSR) at high energies \ci{scm} which can lead to
the CSR  constraints on phenomenological parameters of  a Higgs model.
These constraints on Higgs boson masses (and other parameters) may
serve to pin-point  the signatures of compositeness in the future
experiments on  Higgs particle observation \ci{masha}.

Recently we have developed effective quark models including
higher-dimensional operators made  of fermion fields with
derivatives, which can be used \ci{aa} for the parameterization of
unknown heavy particle dynamics  beyond SM in the spirit of Wilsonian
effective action approach \ci{wk}.  The inclusion of higher
dimensional operators in the fermionic lagrangean of SM opens the
ways to built  the two (and more) Higgs doublet extension  of SM with
composite Higgs bosons.

In this paper we continue the exploration of particular Effective
Quasilocal Quark Models (EQQM) \ci{aary}  which inherit main
properties of a underlying vector gauge theory of QCD type \ci{bark}
(such as technicolor \ci{tec} or topcolor \ci{top} models). The most
important property turns out to be the Chiral Symmetry Breaking (CSB)
at low energies and, on the other hand, the Chiral Symmetry
Restoration (CSR) at high energies. The latter one is controlled by
the Operator Product Expansion of quark current correlators which
include a different number of parity-odd and parity-even
currents. More specifically, here  we deal with the two-point
correlators of scalar and pseudoscalar quark densities \ci{scm,aet}
which are saturated at low energies by Higgs-particle resonances of a
definite parity.  The difference between these correlators is
decreasing rapidly in accordance to OPE of a vector-like gauge theory
\ci{shif,rein}. In the framework of  either EQQM or a low-energy
Higgs-field model it leads to the  CSR constraints on some  parameters
of composite Higgs particles.

We show that CSR at high energies is indeed realized in the 2HQQM of 
type I  in the Nambu-Jona-Lasinio
phase near tricritical point and it is compatible with existence of relatively
light scalar, $h^0$, and pseudoscalar, $A$, Higgs bosons. Thereby 
these models can have their origin  from a QCD-like 
underlying theory with an electroweak compositeness 
scale of order $1 \div 10 TeV$.

Let us remind the effective quark lagrangian of a EQQM which
incorporates all  higher-dimensional operators necessary for the
description of  Two-Higgs Doublet  Standard Model in the low-energy
limit.  The two-flavor, 3d generation quark models with quasilocal
interaction are considered   in which the $ t$- and $ b$-quarks are
involved in the DCSB.

We restrict ourselves  by examination of  the  Two- Higgs 
Quark Models of type I \ci{aary,cve} with Quasilocal interaction (2HQQM)
 with the following
lagrangean: 
\ba 
{\cal L}_{I}&=&\bar q_{L} /\!\!\!\!{\cal D} q_{L} + \bar
t_{R} /\!\!\!\!{\cal D} t_{R}+ \bar b_{R} /\!\!\!\!{\cal D} b_{R}\no && +
\frac{1}{N_{c}\Lambda^{2}}\sum_{k,l=1}^{2}a_{kl}\left( g_{t,
k}J^{T}_{t, k}+g_{b, k}\widetilde J^{T}_{b, k}\right) i\tau_{2}
\left(g_{b, l}J_{b, l}-g_{t, l}\widetilde J_{t, l}
\right). \label{lag} 
\ea 
Here we have introduced the denotations for
doublets of fermion currents: 
\be 
J_{t, k}\equiv \bar
t_{R}f_{t,k}\left( \hat\tau\right)q_{L} ,
\qquad J_{b, k}\equiv \bar b_{R}f_{b,k}
\left(\hat\tau \right)q_{L},\label{J},\qquad
q_{L}=\left( t_{L} \atop b_{L}\right), \la{dou} 
\ee 
and the tilde in
$\widetilde J_{t, k}$  and  $ \widetilde J_{b, k} $ marks charge
conjugated quark currents, rotated with $ \tau_{2} $ Pauli matrix 
\be
\tilde J_{t, k}=i\tau_{2}J^{\star}_{t, k}, \qquad \widetilde J_{b,
k}=i\tau_{2}J^{\star}_{b, k} 
\ee 
The subscripts $ t, b$ indicate right
components of $ t $ and $ b $ quarks in the currents, the index $ k$
enumerates the formfactors: 
\ba && f_{t,1}(\hat\tau) =2-3 \hat\tau ; \quad
f_{t,2}(\hat\tau) = -\sqrt{3} \hat\tau; \qquad \hat\tau \equiv
-\frac{\partial^{2}}{\Lambda^{2}}; \nonumber\\ && f_{b,1}(\hat\tau)=2-
3 \hat\tau,
\quad f_{b,2}(\hat\tau)=-\sqrt{3}\hat\tau ;\label{ff} 
\ea 
which are orthonormal on
the interval $0 \leq \tau = <\hat\tau>\leq 1 $.  In these notations coupling
constants of the four-fermion interaction are represented by $ 2\times
2 $ matrix $ a_{kl} $  and contributed also from the Yukawa constants
$ g_{t, l}, g_{b, k} $.

\section{Effective potential and mass spectrum of composite Higgs bosons}
\hspace*{3ex}
In order to describe the dynamics of composite Higgs bosons
the lagrangean density (\ref{lag}) of the Model I 
must be rearranged by means of
introduction of auxiliary bosonic variables and by integrating out
fermionic degrees of freedom \ci{aa}.  Namely, we
define two scalar $SU(2)_{L}$-isodoublets:
\be
 \Phi_{1}=\left(\phi_{11}\atop \phi_{12}\right),\qquad
  \Phi_{2}=\left(\phi_{21}\atop \phi_{22}\right)
\ee
and their charge conjugates:
\be
  \widetilde \Phi_{1}=
\left(\phi_{12}^{\star}\atop -\phi_{11}^{\star}\right),\qquad
  \widetilde \Phi_{2}=
\left(\phi_{22}^{\star}\atop -\phi_{21}^{\star}\right).
\ee
Then the lagrangean (\ref{lag}) can be
rewritten in the following way:
\be
{\cal L}_{I} =  L_{kin} +
N_{c}\Lambda^{2}\sum_{k,l=1}^{2}\Phi^{\dagger}_{k}
(a^{-1})_{kl}\Phi_{l}+i \sum_{k=1}^{2}\left[
g_{t, k}\widetilde \Phi_{k}^{\dagger}J_{t, k}+
g_{b, k}\Phi_{k}^{\dagger}J_{b, k}\right] + h.c. \la{yuka}
\ee
The integrating out of  fermionic degrees of freedom will produce
the effective action for Higgs bosons of which we shall keep  only the
kinetic term and  the effective potential
consisting of two- and four-particles
operators.
The omitted terms are supposedly small,
being proportional to inverse powers of 
a large scale factor $\Lambda$ and/or of a large
$\ln(\Lambda/v);\, v \simeq 246 GeV$.
The Yukawa constants are chosen of the form
\be
 g_{t,k}=1 ; \qquad g_{b,k}=g
\ee
for $k= 1,2$. The first choice can be done because the
fields
$ \Phi_{1} $ and $ \Phi_{2} $ can always be rescaled by an arbitrary
factor which is absorbed by 
redefinition of polycritical coupling constants 
$a_{kl}$. Other constants $g_{b,k}$
are taken equal for the simplicity. Their value
$ g $ induces the quark mass ratio
$ m_{b}/m_{t} $.

A systematic approximation can be developed in the vicinity of (poly)critical
point,
\be
8\pi^2 a_{kl}^{-1}\sim\delta_{kl}+\frac{\Delta_{kl}}{\Lambda^2}, \qquad
      |\Delta_{kl}|\ll \Lambda^2 ,                           \label{11}
\ee
which signifies the cancellation of quadratic divergences \ci{aay}. 

Then the effective potential of Higgs fields in 2HQQM of type I  reads:
\ba
V_{eff}&=&
\frac{N_{c}}{8\pi^2}\Biggl(
-\sum_{k,l=1}^{2}(\Phi^{\dagger}_{k}\Phi_{l})\Delta_{kl} - 
8g^4(\Phi^{\dagger}_{1}\Phi_{1})^{2}\ln g^2 \no
&&+ (1 + g^4)\left[
8(\Phi^{\dagger}_{1}\Phi_{1})^{2}\left(
\ln\frac{\Lambda^{2}}{4(\Phi^{\dagger}_{1}\Phi_{1})}+\frac12\right)+
\right.\no
&&\left.-\frac{159}{8}(\Phi^{\dagger}_{1}\Phi_{1})^{2}
+\frac{9}{8}(\Phi^{\dagger}_{2}\Phi_{2})^{2}
+\frac{3}{4}(\Phi^{\dagger}_{1}\Phi_{1})(\Phi^{\dagger}_{2}\Phi_{2})+
\right.\nonumber\\
&&\left.+\frac{3}{4}(\Phi^{\dagger}_{1}\Phi_{2})(\Phi^{\dagger}_{2}\Phi_{1})+
\frac{3}{8} (\Phi^{\dagger}_{1}\Phi_{2})^{2}+
\frac{3}{8} (\Phi^{\dagger}_{2}\Phi_{1})^{2}-
\right.\nonumber\\
&&\left.-\frac{5\sqrt{3}}{4}
(\Phi^{\dagger}_{1}\Phi_{1})\left((\Phi^{\dagger}_{1}\Phi_{2})+
(\Phi^{\dagger}_{2}\Phi_{1})\right)+\right.\nonumber\\
&&\left.+\frac{\sqrt{3}}{4}(\Phi^{\dagger}_{2}\Phi_{2})
\left((\Phi^{\dagger}_{1}\Phi_{2})+(\Phi^{\dagger}_{2}\Phi_{1})\right)
\right]\Biggr)+O\left(\frac{\ln\Lambda}{\Lambda^{2}}\right),\label{Vmod1}
\end{eqnarray}
where the bilinear ``mass'' term is in general non-diagonal  and  represented
by the real, symmetric
$ 2\times 2 $ matrix
$ \Delta_{kl} $. This is the two-Higgs potential in the large $N_c$ approach and
a more realistic potential should involve the true Renormalization Group flow
of Two-Higgs Doublet SM \footnote{see the updated one-loop RG equations in \ci{cvek}
and references therein.} with initial conditions at high energies taken from \gl{Vmod1}. 

In general there exist  the regimes where at the minimum of the effective potential
the ratio of v.e.v. of neutral Higgs fields  
is complex \ci{aay} and v.e.v. of charged components are also non-zero.  Due to $U(2)$
symmetry of the effective potential \gl{Vmod1} one can always choose the parameterization
of Higgs fields in the vicinity of a minimum with  only one of v.e.v. being complex:
\be
\langle\Phi_{1}\rangle = \frac{1}{\sqrt2}\left( 0 \atop \phi_1\right),\qquad
 \langle\Phi_{2}\rangle = \frac{1}{\sqrt2}\left(\theta \atop \phi_2 + i \rho\right) . \la{vev}
\ee 
The appearance of complex a v.e.v. leads  to CP violation \ci{lee,br,osl,bern} and
a non-zero v.e.v. of charged fields breaks electric stability of the vacuum and supplies the photon with a mass. 
The latter phase is not realized in the Universe at zero temperatures as it follows from the severe bounds on
both electric charge asymmetry and on the photon mass  (see, e.g. \ci{lah}).
 
The condition  of minimum of the potential (\ref{Vmod1})
brings the mass-gap equations for them
which solution may cause the DCSB if it is an absolute  minimum.
In the explicit form they are:
\begin{eqnarray}
\frac{8\pi^2}{N_c} \frac{\delta V_{eff}(\phi_k, \rho, \theta)}{\delta  \phi_1} &=&
-\Delta_{11} \phi_1 -  \Delta_{12} \phi_2 + 8 \phi_1^{3}\ln\frac{\Lambda^{2}}{2 \phi_1^{2}}
+8 \phi_1^{3}g^{4}\ln\frac{\Lambda^{2}}{2g^{2} \phi_1^{2}}\\
&&+\frac{1+g^{4}}{8}\Bigl[- 159 \phi_1^{3}
- 15\sqrt{3} \phi_1^2 \phi_2
+ 9 \phi_1 \phi_2^{2}+ \sqrt{3} \phi_2^{3}\no
&&+ \sqrt{3} (\theta^2 + \rho^2) (\sqrt3 \phi_1 + \phi_2) \Bigr] = 0,\no
\frac{8\pi^2}{N_c} \frac{\delta V_{eff}(\phi_k, \rho, \theta)}{\delta  \phi_2} &=&
-\Delta_{12} \phi_1 -  \Delta_{22} \phi_2 + \frac{1+g^{4}}{8}\Bigl[- 5\sqrt{3} \phi_1^{3}
+ 9 \phi_1^{2} \phi_2\no
&&+3\sqrt{3} \phi_1 \phi_2^2
+ 9 \phi_2^{3} + \sqrt3  (\theta^2 + \rho^2) (\phi_1 + 3\sqrt3 \phi_2) \Bigr] = 0\no 
\frac{8\pi^2}{N_c} \frac{\delta V_{eff}(\phi_k, \rho, \theta)}{\delta  \theta} &=&
\theta \left[ -  \Delta_{22}  + \frac{1+g^{4}}{8}\left( 9 (\theta^2 + \rho^2) +
(\sqrt3 \phi_1  + \phi_2)^2 +  8\phi_2^2 \right)\right]
= 0;\no
\frac{8\pi^2}{N_c} \frac{\delta V_{eff}(\phi_k, \rho, \theta)}{\delta  \rho} &=&
\rho \left[ -  \Delta_{22}  + \frac{1+g^{4}}{8}\left( 9 (\theta^2 + \rho^2) + (\sqrt3 \phi_1  + \phi_2)^2 +  
8\phi_2^2\right)\right] 
= 0 \nonumber
\la{mg}
\end{eqnarray}
  We assume the electric charge stability  of the vacuum, {\it i.e.}
that only  neutral components  of both Higgs doublets
may have nonzero v.e.v. 
\ci{gun,krod,rod}. Evidently, the relevant solutions with $\theta = \rho =0$ are unique if
\be
\Delta_{22}   \leq  (1+g^{4} )\left[ \phi_2^2 + \frac18 (\sqrt3 \phi_1 + \phi_2)^2\right] .
\ee
 This  bound follows also from the absence of tachyons among composite
bosons which will be derived later on. On the other hand, 
due to $SO(2)$ symmetry of effective potential \gl{Vmod1} under 
rotations of $\theta$ and $\rho$,   typically, 
the CP violating  solutions for 2HQQM of 
type I break also the electroneutrality
of the vacuum unless a special fine-tuning is performed. 
It gives us one more argument against the realization
of CP-violating phase in such models.

One can choose $\phi_1, \phi_2$ as independent scales of  2HQQM. Then the mass-gap equations \gl{mg} allow to
find the parameters $\Delta_{11}, \Delta_{12}$ describing the deviation from critical coupling constants  
as functions of   $\phi_1, \phi_2, \Delta_{22} $.

The true minimum  is derived from the positivity of the second variation
of the effective action around a solution of the mass-gap equation,
\be
\Phi_{k} = \left( \pi_k^+ \atop \frac{1}{\sqrt2}(\phi_k + \sigma_k + i \pi_k)\right) ;\, k = 1,2 .
\ee
This variation reads:
\ba
\frac{16\pi^2}{N_c}  S_{eff}^{(2)} &\equiv&
\frac12 \left(\sigma,(\hat A^{\sigma} p^2 + \hat B^{\sigma})
\sigma\right)\no
&&+ \left(\pi^-,(\hat A^{\pi^+} p^2 +
\hat B^{\pi^+}) \pi^+\right) + \frac12 \left(\pi,(\hat A^{\pi} p^2 +
\hat B^{\pi}) \pi\right), \la{qua}
\ea
where two symmetric matrices - for
the kinetic term $\hat A^i = \left(A^{i}_{kl}\right),\quad i =(\sigma,\pi^+, \pi)$ 
and for the constant, momentum independent part,
$\hat B^i = \left(B^{i}_{kl}\right) $ -  have been introduced
in the CP conserving phase.

The mass spectrum of related bosonic states is
determined by the solutions of the secular equations :
\be
\det(\hat A^{i} p^{2}+\hat B^{i})=0, \label{spectrumeqn}
\ee
at $-m^2 = p^2 < 0$ in both scalar and pseudoscalar channels.

The kinetic matrix $\hat{A}$ as being multiplied by $p^{2}$ is derived in the
soft-momentum expansion in powers of $p^{2}$:
\be
{\cal L}_{\mbox{kin}} =  \frac{N_{c}}{16\pi^2}
        \sum_{k,l=1}^{2}
        \Biggl(I_{kl}^{(1)}\partial_{\mu}
 \pi^-_k \,\partial_{\mu}
\pi^+_l +
        \frac12  I_{kl}^{(2)}\left(\partial_{\mu}
 \sigma_k  \partial_{\mu}
 \sigma_l + \partial_{\mu}
 \pi_k  \partial_{\mu}
 \pi_l \right)\Biggr) \label{sum1}
\ee
where $ I^{(1)}_{kl} $ contributes to the kinetic term for charged components
of Higgs doublets and $ I^{(2)}_{kl} $ defines the latter one 
for the neutral components:
\ba
  A_{kl}^{\sigma}
 &\simeq&  A_{kl}^{\pi^+}   \simeq  A_{kl}^{\pi}    \simeq   
I_{kl}^{(1)} \simeq    I_{kl}^{(2)} \no
&& \simeq 
           f_{t,k}(0) f_{t,l}(0) 
\left(\ln\frac{\Lambda^2}{m_t^{2}} - 1\right) + g^2 f_{b,k}(0) 
f_{b,l}(0) \left(\ln\frac{\Lambda^2}{m_b^{2}} - 1\right)
        \nonumber \\
        &&+\int\limits_{0}^{1}\bigl(f_{t,k}(\tau) f_{t,l}(\tau)
                + g^2 f_{b,k}(\tau) f_{b,l}(\tau)-\nonumber \\
       &&- f_{t,k}(0) f_{t,l}(0)- g^2 f_{b,k}(0) f_{b,l}(0)
        \bigr)\frac{d\tau}{\tau} + O\left(\frac{\ln
\frac{\Lambda^{2}}{m_t^2}}{\Lambda^{2}}\right).
\ea
 The related integrals for  $I^{(1),(2)}_{kl}$ have been calculated at large values
of
$ \Lambda $, 
in the large-log limit $\ln(\Lambda^2/m^2_t)\simeq\ln(\Lambda^2/m^2_b) \gg 1 $
and for the CP conserving phase. 
Further on we consider  $g \ll 1$ as $m_b \ll m_t$. As well in the matrix elements containing large 
$\ln(\Lambda^2/m_t^2) $ we neglect  isospin breaking effects. 
After substitution of \gl{ff} kinetic matrices
$\hat A$ take the form:
\be
 A_{kl}^{\sigma}
 \simeq A_{kl}^{\pi}\simeq  A_{kl}^{\pi^+} \simeq \left(\begin{array}{cc}
\left(4\ln\frac{\Lambda^2}{m^2_t} - \frac{23}{2}\right) &\quad - \frac{\sqrt{3}}{2}\\
- \frac{\sqrt{3}}{2} & \quad \frac32
\end{array}\right) \la{mA}
\ee

Let us now obtain the matrix of second variations $\hat B^{i}$ of the effective potential
for the  Model I :
\ba
B_{kl}^{\sigma}&=&\frac{16\pi^2}{N_{c}}\frac{\partial^{2}}{\partial\sigma_{k}
\partial\sigma_{l}}V_{eff} ,
\qquad B_{kl}^{\pi}=\frac{16\pi^2}{N_{c}}\frac{\partial^{2}}{\partial\pi_{k}
\partial\pi_{l}}V_{eff} ,\no
B_{kl}^{\pi^+} &=&\frac{8\pi^2}{N_{c}}\frac{\partial^{2}}{\partial\pi^+_{k}
\partial\pi^-_{l}}V_{eff} ,\quad \mbox{for} \quad
\Phi_k = \langle\Phi_k\rangle .
\ea
In the above approximation:  $B_{kl}^{\pi^+} \simeq  B_{kl}^{\pi}$ 
and therefore the spectra of charged
and neutral pseudoscalars coincide. Other elements of matrices  $\hat B$ are:
\begin{eqnarray}
B_{11}^{\sigma}&=&
        - 2\Delta_{11}  + 48\phi_{1}^{2}
        \ln\left({{\Lambda^{2}}\over{m_t^2}}\right)
        -\frac{605}{4}\phi_{1}^{2}-\frac{15\sqrt{3}}{2}\phi_{1}\phi_{2}
        + \frac{9}{4} \phi_{2}^{2},\no   
B_{12}^{\sigma}&=& -2\Delta_{12} -\frac{15\sqrt{3}}{4}\phi_{1}^{2} +
        \frac92 \phi_{1}\phi_{2}+
        \frac{3\sqrt{3}}{4}\phi_{2}^{2} ,\no
B_{22}^{\sigma}&=& -2\Delta_{22} +
        \frac94 \phi_{1}^{2}+ \frac{3\sqrt{3}}{2}\phi_{1}\phi_{2} +\frac{27}{4} \phi_{2}^{2},\no
B_{11}^{\pi}&=&        - 2\Delta_{11}  + 16\phi_{1}^{2}
        \ln\left({{\Lambda^{2}}\over{m_t^2}}\right)
        -\frac{159}{4}\phi_{1}^{2}-\frac{5\sqrt{3}}{2}\phi_{1}\phi_{2}
        + \frac{3}{4} \phi_{2}^{2},\no   
B_{12}^{\pi}&=&-2\Delta_{12}
        - \frac{5\sqrt{3}}{4}\phi_{1}^{2}+ \frac32\phi_{1}\phi_{2} +\frac{\sqrt{3}}{4}\phi_{2}^{2}, \no
B_{22}^{\pi} &=&-2\Delta_{22} +
        \frac34 \phi_{1}^{2}+ \frac{\sqrt{3}}{2}\phi_{1}\phi_{2} +\frac{9}{4} \phi_{2}^{2} . \la{mB}
\end{eqnarray}
For large $\ln\left(\Lambda^{2}/m_t^2\right) \gg 1$ both the solutions of \gl{mg} 
and the mass spectrum look differently in two regimes: $ 
\Delta_{11} \sim m^2_t\ln\left(\Lambda^{2}/m_t^2\right) \gg 
\Delta_{12} \sim \Delta_{22} \sim m^2_t$  or  $ \Delta_{11} \sim 
\Delta_{12} \sim \Delta_{22} \sim  m^2_t\ln\left(\Lambda^{2}/m_t^2\right) \gg m^2_t$. 

In the first case the mass spectrum of physical scalar bosons 
$H^0, h^0$ reads:
\ba
&& m^2_{H^0} = 8 \phi_1^2 = 4 m_t^2 \simeq \frac{\Delta_{11}}{\ln\left(\Lambda^{2}/m_t^2\right)};\no
&& m^2_{h^0}  = \frac12\left(8 \phi_2^2  + (\sqrt3 \phi_1 +\phi_2)^2\right) - \frac43 \Delta_{22} \equiv
\frac32 \delta m^2 - \frac43 \Delta_{22},
\la{mas1}
\ea
and for physical pseudoscalar bosons $\pi, A, H^\pm $:
\be
m^2_\pi = 0;\qquad m^2_{A} =  m^2_{ H^\pm} =\frac16\left(8 \phi_2^2  + (\sqrt3 \phi_1 +\phi_2)^2\right) - 
\frac43 \Delta_{22} \equiv
\frac12 \delta m^2 - \frac43 \Delta_{22}, \la{mas2}
\ee
where we have introduced the notation for the mass splitting between excited states:
\be
 m^2_{h^0}  - m^2_{ A} = \delta m^2. \la{split}
\ee
One can see that, in general, $ m^2_{h^0}  - m^2_{A} = \delta m^2 \not = m^2_{H^0}  - m^2_\pi $ 
and the splitting between scalar and pseudoscalar states is not equidistant. But if the dynamical mass is a true constant,
{\it i.e.} $ \sqrt3 \phi_1 + \phi_2 = 0$, then
 one recovers the remarkable relation:
\be
  m^2_{h^0}  - m^2_{A}  = m^2_{H^0}  - m^2_\pi = 4 m^2_t .  
\la{remark}
\ee
As $ m^2_t = 2 \phi_1^2$  this parameter is fixed from the phenomenology and never takes zero values.
Then for a given $\Delta_{22}$ the lowest masses of excited bosons arise at $\phi_1 = - 3\sqrt3 \phi_2 $ and
amount to:
\ba
&&\delta m^2_{min} = \frac19 m^2_{H^0};\qquad  m^2_{h^0} = \frac16 m^2_ {H^0}- \frac43 \Delta_{22} 
\geq \frac19 m^2_{H^0};\no
&&  m^2_ A= \frac{1}{18} m^2_ {H^0}- \frac43 \Delta_{22} \geq 0. \la{smin}
\ea 
It follows from these inequalities that the second pseudoscalar Higgs boson can in principle be very light even massless
whereas the second scalar Higgs particle can be lighter than the first one (of Nambu-Jona-Lasinio type) but still enough
heavy. 

For instance, if we treat these relations at 
the tree level then: $m_{H^0}= 2 m_t \simeq 350 GeV$ but
the lightest mass for the second Higgs boson  
$ m^2_{h^0} =  \frac19 m^2_{H^0} + m^2_A \geq
130 GeV$ for  $  m_A \geq 50 GeV$ (see \ci{pdg}).

In the case when $m_t^2\ln\left(\Lambda^{2}/m_t^2\right) \sim \Delta_{11} \sim 
\Delta_{12} \sim \Delta_{22} \gg m^2_t$ the mass spectra can be easily derived from 
Eqs.\gl{mas1}, \gl{mas2} retaining the last term, $ m^2_{h^0}  \simeq  m^2_ A \simeq - \frac43 \Delta_{22}$ and
their splitting is again described by  \gl{split}. Thus for such parameters  $\Delta_{ik}$ the second multiplet of
Higgs boson is indeed much heavier 
that the first one $  m^2_ {H^0} \ll  m^2_{h^0}  \simeq  m^2_ A \sim 
m_t^2 \ln\left(\Lambda^{2}/m_t^2\right) \sim 1 (TeV)^2$. 
Thus this scenario is  
practically equivalent to a One-Higgs SM for accessible energies. Therefore we will not consider it further on.

\section{Chiral symmetry restoration in QCD-like models}
In large $N_c$ QCD-like models \ci{nc,wit} the leading contributions into
two-point correlators of scalar and pseudoscalar
quark densities are given by sums
over an  (infinite) number of meson poles: 
\ba
\Pi_\sigma (p^2) &=&- \int d^4x \,\exp(ipx)\
\langle T\left(\bar q q (x) \,\, \bar q q(0)\right)\rangle \no
&=&
\sum_n \,\frac{Z^\sigma_n}{p^2 + m^2_{\sigma,n}} + C^\sigma_0 + C^\sigma_1 p^2,\no
\Pi_\pi^{ab} (p^2)&=& \int d^4x \,\exp(ipx)\
\langle T\left(\bar q\gamma_5 \tau^a q (x) \,\,
\bar q\gamma_5 \tau^b q (0)\right)\rangle \no
 &=&  \delta^{ab}\left(
\sum_n \,\frac{Z^\pi_n}{p^2 + m^2_{\pi,n}}  + C^\pi_0 + C^\pi_1 p^2\right).
\la{planar}
\ea
$C_0$ and $C_1$ are contact terms required for the regularization of infinite sums.

The high-energy asymptotics are given by 
perturbation theory and operator product expansion taking into account
the asymptotic freedom of QCD-like interaction. As well the nonperturbative
generation of (techni) gluon and quark condensates \ci{shif} is 
assumed to determine subleading power-like
corrections to perturbative asymptotics.

The covariant derivative in a QCD-like gauge theory,
$i\partial_{\mu}\rightarrow iD_{\mu}
=i\partial_{\mu} + G_{\mu}$, contains gluon fields
$G_{\mu} \equiv  g_s \lambda^a G^a_{\mu}$,
where $tr(\lambda^a \lambda^b) = 2 \delta ^{ab}$. 
The related gluon field strength is defined as
$G_{\mu\nu} \equiv - i [D_{\mu}, D_{\nu}]$.  $g_s$ is 
the gauge coupling constant. Then the conventional QCD coupling constant is 
$\alpha_s \equiv g_s /4\pi$.  Respectively,
 in the chiral limit ($m_q = 0$)
the scalar and pseudoscalar
 correlators have the following 
behavior \ci{shif,rein,jam} at large $p^2$ 
(for one-flavor quarks) motivated by 
Operator Product Expansion:
\ba
\Pi_{\sigma,\pi}(p^2) |_{p^2 \rightarrow \infty} 
&\simeq & \frac{N_c}{8\pi^2}\left(
1 + \frac{11 N_c \alpha_s}{8 \pi}\right) p^2\,\ln\frac{p^2}{\mu^2}
\no
&&+ \frac{\alpha_s}{8\pi p^2} \langle (G^a_{\mu\nu})^2\rangle
+ \frac{\pi \alpha_s}{3 p^4}
 \langle \bar q \gamma_{\mu} \lambda^k q
\bar q \gamma_{\mu} \lambda^k q\rangle \no
&&
\mp \frac{\pi\alpha_s}{2p^4} \langle \bar q \sigma_{\mu\nu} \lambda^k q
\bar q \sigma_{\mu\nu} \lambda^k q\rangle +
{\cal O}(\frac{1}{p^6}), \la{as}
\ea
in Euclidean notations. 
Herein it is assumed that $\alpha_s (\mu)\ll 1;\quad \mu > \Lambda_{CSB}$. 
In the large-$N_c$ limit:
\be
\left(\Pi_P(p^2)- \Pi_S(p^2)\right)_{p^2 \rightarrow \infty} \equiv
\frac{\Delta_{SP}}{p^4}  + 
{\cal O} \left(\frac{1}{p^6}\right);\quad
\Delta_{SP} \simeq  12 \pi\alpha_s \left(<\bar q q>\right)^2,   \la{CSR}
\ee
where the vacuum dominance hypothesis \ci{shif}  has been exploited
to estimate four-quark condensate contribution in terms of bilinear
quark condensates $<\bar q q>$.
This rapidly decreasing asymptotics is a consequence of the
chiral invariance of the lagrangean of a massless QCD-like theory.

When comparing \gl{planar} and  the first term of \gl{as} one can see that
in order to reproduce
the perturbative asymptotics
the  infinite number of
resonances with the same quantum numbers should exist in each channel. 

On the other hand,
in the difference of scalar and pseudoscalar correlators the saturation may be successfully delivered
by a finite number of low-lying resonances due to  CSR \ci{scm, aet}. 

Thus CSR
at high energies may be thought of  as a possible constraint
on the EQQM to be a manifestation of compositeness in the Higgs model
 \ci{aay,rk}. We shall demand that, at the compositeness scale $\Lambda_{CSB}$,
the relation \gl{CSR} is approximately fulfilled. 

Following the planar limit of the QCD-like interaction, eq.\gl{planar} one can
make the
two-resonance ansatz for scalar and pseudoscalar correlators provided by
Two-channel EQQM model:
\ba
\Pi_\sigma (p) &=& 
\frac{Z_{\sigma}}{p^2 + m^2_{\sigma}} + \frac{Z_{\sigma'}}{p^2
+ m^2_{\sigma'}} + C^{\sigma}_0;\no
\Pi_\pi (p) &=& 
\frac{Z_{\pi}}{p^2}+
\frac{Z_{\pi'}}{p^2 + m^2_{\pi'}}  + C^{\pi}_0. \la{ans}
\ea
We remark that for this type of  models the constants can be taken 
$C_1^{\sigma}= C_1^{\pi} = 0$.
 From the requirement of asymptotic CSR (\ref{CSR}) it follows
that:
\ba
C^{\sigma}_0 &=& C^{\pi}_0 \equiv C \left(= \frac{<\bar q q>}{m_{dyn}} < 0\right);\\
Z_{\sigma} + Z_{\sigma'} &=& Z_{\pi} + Z_{\pi'};\qquad
Z_{\sigma} m^2_{\sigma} + Z_{\sigma'} m^2_{\sigma'}
=  Z_{\pi'} m^2_{\pi'} + \Delta_{SP} . \la{constr}
\ea
The first two relations can be fulfilled in the conventional NJL model
which corresponds to the one-resonance ansatz, $Z_{\sigma',\pi'} = 0$,
whereas the last one can be provided only in a two-resonance model,
for the  $\Delta _{SP}$ in \gl{CSR} (see below). 

In order to apply the above CSR sum rules we have to obtain the appropriate two-point correlators
for scalar and pseudoscalar composite Higgs fields.
\section{CSR sum rules in 2HQQM}
\hspace*{3ex} Let us introduce in 2HQQM the  
external sources which couple to the scalar and pseudoscalar quark densities. With their help one
can easily  derive  required quark correlators. As these densities
 are doublets in our model , eq.\gl{dou}, the relevant sources 
$X_1 =(\chi_{1j}), X_{2} = (\chi_{2j})$ are taken as doublets as well.
The sources are complex: $
\chi_{kj} = S_{kj} + i P_{kj},$
generating both scalar and pseudoscalar densities.
The structure of the  corresponding operators in the quark lagrangean is designed  similar to
the Yukawa operators in \gl{yuka}:
\be
{\cal L}_{I} (S, P) = 
i \sum_{k=1}^{2}\left[
g_{t, k}\widetilde X_{k}^{\dagger}J_{t, k}+
g_{b, k} X_{k}^{\dagger}J_{b, k}\right] + h.c. \la{yukasp}
\ee
We remind that in 2HQQM under consideration: $g_{t, k} =1 \gg g_{b, k} = g$ and take for simplicity
$g= 0$.
Then  the effect of external sources can be separated by shifting the Higgs fields
in the quark part of the lagrangean  \gl{yuka}:
\be
\phi_{kj} \longrightarrow  \bar\phi_{kj} + \chi_{kj}.
\ee
The  dynamical boson fields arise as fluctuations around the solutions of the mass-gap
equation \gl{mg}:
\be
 \bar\phi_{k1} = \phi_{k1} -  \chi_{k1};\quad \bar\phi_{k2} = \phi_{k2} -  \chi_{k2}
 + <\bar\phi_{k2}>.
\ee
In terms of doublets $X_k,\, \Phi_k$ the supplementary lagrangean takes the form:
\ba
\Delta {\cal L}_I &=& N_{c}\Lambda^{2}\sum_{k,l=1}^{2}\left[X^{\dagger}_{k} (a^{-1})_{kl}X_{l}- 
X^{\dagger}_{k} (a^{-1})_{kl}\Phi_{l}\right.\no
&&\left. -
\Phi^{\dagger}_{k} (a^{-1})_{kl}X_{l} - 
X^{\dagger}_{k} (a^{-1})_{kl} \langle\Phi_{l}\rangle 
-  \langle\Phi^{\dagger}_{k}\rangle (a^{-1})_{kl}X_{l}\right]. \la{qua1}
\ea
The terms linear in
external sources are irrelevant for correlators and will be neglected further on.
As we study the CP conserving phase with non-trivial v.e.v. for scalar fields only
the physical parameters of neutral and charged pseudoscalar bosons are the same. Therefore
we can restrict ourselves with neutral components only.   Then with notations:
\be
\phi_{k2} = \frac{1}{\sqrt2}(\sigma_k + i \pi_k);\qquad \chi_{k2} =  \frac{1}{\sqrt2}(S_k + i \,P_k),
\ee 
one obtains the quadratic part of the lagrangean ${\cal L}_I$ consisted of \gl{qua}
and  \gl{qua1}. As it is gaussian one can easily unravel the dependence on external
fields with the help of classical Eqs. of motion:
\ba
\sigma_k^{\mbox{cl}} &=& 16 \pi^2 
\Lambda^2\sum_{l,m=1}^{2} 
\left( A^{\sigma\sigma} p^2 + B^{\sigma\sigma} \right)_{kl}^{-1}a^{-1}_{lm} S_m \simeq 
2\Lambda^2\sum_{l=1}^{2}
\left( A^{\sigma\sigma} p^2 + B^{\sigma\sigma} \right)_{kl}^{-1} S_l \no
\pi_k^{\mbox{cl}} &=& 16 \pi^2 \Lambda^2\sum_{l,m=1}^{2} 
\left( A^{\pi\pi} p^2 + B^{\pi\pi}\right)_{kl}^{-1} a^{-1}_{lm} P_m \simeq 
2\Lambda^2\sum_{l=1}^{2} 
\left( A^{\pi\pi} p^2 + B^{\pi\pi}\right)_{kl}^{-1}  P_l ,
\ea
which is simplified in the vicinity of polycritical point, \quad $ 8 \pi^2 a^{-1}_{kl} \simeq \delta_{kl}$.

The resulting effective action for generating of two-point correlators is given by:
\ba
S^{(2)}& \simeq& \frac{N_c\Lambda^2}{16\pi^2} \sum_{k,l=1}^{2}\left(
S_k \Gamma^{(\sigma)}_{kl} S_l +  P_k \Gamma^{(\pi)}_{kl}P_l\right)\no
 \Gamma^{(\sigma)}_{kl}& =&\delta_{kl} - 2\Lambda^2 \left(A^{\sigma\sigma} p^2 +
 B^{\sigma\sigma}\right)^{-1}_{kl};\quad
\Gamma^{(\pi)}_{kl} =\delta_{kl}  - 2\Lambda^2 \left(A^{\pi\pi} p^2 +
 B^{\pi\pi}\right)^{-1}_{kl}. \la{cor1}
\ea
When taking the approximate expressions \gl{mA}, \gl{mB} for matrices $\hat A, \hat B$
one derives the inverse propagators.
In particular, the strictly local quark density can be presented as a superposition of
two currents:
\be
\bar t t = \frac12 (\bar t f_1 t - \sqrt{3} \bar t f_2 t),
\ee 
and in the scalar channel the related correlator \gl{planar}
has the following form:
\ba
\Pi_\sigma (p^2) &=& - \frac{N_c \Lambda^2}{16\pi^2} \left[\Gamma^{(\sigma)}_{11} +
3\Gamma^{(\sigma)}_{22} - 2\sqrt{3}\Gamma^{(\sigma)}_{12}\right] 
= C^\sigma + \frac{Z_{H^0}}{p^2 +m^2_{H^0}} + \frac{Z_{h^0}}{p^2 +m^2_{h^0}} ;\no 
C^\sigma &=& - \frac{N_c\Lambda^2}{4\pi^2} \simeq \frac{<\bar t t>}{m_t};\no
Z_{H^0} &=& \frac{N_c\Lambda^4}{48\pi^2\left(\ln\frac{\Lambda^2}{m^2_t} 
- 3\right)\,
(m^2_{H^0} -m^2_{h^0})}\no
&&\times \left[ 12 \left(\ln\frac{\Lambda^2}{m^2_t} 
- 3\right) m^2_{H^0} +  6\Delta_{11} +4\sqrt{3} \Delta_{12} + 
 2\Delta_{22}  - 144  \phi_1^2 \ln\frac{\Lambda^2}{m^2_t} +
 474 \phi_1^2 \right.\no
&&\left. + 12\sqrt3 \phi_1\phi_2 - 18 \phi_2^2  \right] 
= O\left(\frac{1}{\ln\frac{\Lambda^2}{m^2_t}}\right);\no
Z_{h^0} &=& -  Z_{H^0} + \frac{N_c\Lambda^4}{4\pi^2}; \la{con1}
\ea
and similarly in the pseudoscalar channel,
\ba
\Pi_\pi (p^2) &=& - \frac{N_c \Lambda^2}{16\pi^2} \left[\Gamma^{(\pi)}_{11} +
3\Gamma^{(\pi)}_{22} - 2\sqrt{3}\Gamma^{(\pi)}_{12}\right] 
= C^\pi + \frac{Z^\pi}{p^2} + \frac{Z_{A}}{p^2 +m^2_{A}} ;\no 
C^\pi &=& C^\sigma = - \frac{N_c\Lambda^2}{4\pi^2} \simeq \frac{<\bar t t>}{m_t};\no
Z^\pi &=& - \frac{N_c\Lambda^4}{48\pi^2\left(\ln\frac{\Lambda^2}{m^2_t} 
- 3\right)\,
m^2_{A} } \left[2\Delta_{22} + \Delta_{12}\left( 4\sqrt{3} - 6 \frac{\phi_2}{\phi_1}\right)
+  \frac{27}{4} \phi_1^2  - \frac{29\sqrt3}{4} \phi_1\phi_2\right.\no
&&\left.+ \frac{3}{4} \phi_2^2  + \frac{3\sqrt3}{4}   \frac{\phi_2^3}{\phi_1}\right] 
= O\left(\frac{1}{\ln\frac{\Lambda^2}{m^2_t}}\right);\no
Z_{A} &=&  -  Z_{\pi} + \frac{N_c\Lambda^4}{4\pi^2}. \la{con2}
\ea
We stress that the residues in poles are of different order of magnitude: 
\be
Z_{H^0} \sim Z^\pi \sim {1 \over \ln\frac{\Lambda^2}{m^2_t}} \ll Z_{h^0} \sim Z_{A}, \la{res}
\ee 
which can be seen from the first Eq. in \gl{con1} after using the mass-gap Eqs.\gl{mg} 
and taking the value \gl{mas1} of
the scalar boson mass, $m^2_{H^0} \simeq 8\phi_1^2$. Indeed then
the logarithms cancel each other in the combination,  $12 m^2_{H^0} \ln\frac{\Lambda^2}{m^2_t} 
m^2_{H^0} +  6\Delta_{11} - 144  \phi_1^2 \ln\frac{\Lambda^2}{m^2_t}$.

Now we are able to impose and check the CSR constraints \gl{constr}.
First of all one can see that the chiral symmetry is not broken in leading asymptotics,
$C^\pi = C^\sigma $. Next we check the subleading asymptotics which represents the
generalized $\sigma$-model relation:
\be
Z_{H^0} + Z_{h^0} = Z_{\pi} + Z_{A} = \frac{N_c\Lambda^4}{4\pi^2}, \la{zeta}
\ee
which is fulfilled, in fact, at a high precision including subleading  $ 1/\ln\Lambda^2$ terms.

The last constraint in \gl{constr} can be satisfied for an  appropriate 4-fermion condensate
contribution $\Delta_{SP}$. Thus:
\be
Z_{H^0} m^2_{H^0} + Z_{h^0} m^2_{h^0} =  Z_{A} m^2_{A} + \Delta_{SP}.
\la{last}
\ee
Taking into account \gl{res} and the sum rule \gl{zeta} one concludes that:
\be
Z_{h^0} \simeq Z_{A} \simeq \frac{N_c\Lambda^4}{4\pi^2};\qquad 
\frac{N_c\Lambda^4}{4\pi^2}\left(m^2_{h^0} - m^2_{A}\right) = \Delta_{SP}.
\ee
In order to estimate a maximal scale of compositeness  
$\Lambda \equiv \Lambda_{C}$ due to QCD-like forces we assume that presumably
the typical (techni) QCD scale is of order $\Lambda_{TQCD} \sim m_t \ll 
\Lambda_{C} $ and 
one can apply effectively the one-loop approximation to calculate the strong interaction constant $\alpha_s$ (see, \ci{pdg}).
As it follows from \gl{smin}, the minimal value of mass splitting corresponds to $m^2_{H^0}/9$. Respectively the value
of $\Delta_{SP}$ is described by the relation \gl{CSR} and in the present 2HQQM the condensates are given in \gl{con1}. 
Let us combine these estimates and relations to derive the upper bound on the compositeness scale  $\Lambda_{C}$:
\be
 \frac{N_c\Lambda_C^4 m^2_t}{9\pi^2} =  \mbox{\rm min}(\Delta_{SP}) = 
12\pi\alpha_s \frac{N^2_c\Lambda_C^4 m^2_t}{16\pi^4};
\ee
hence:
\be
\alpha_s N_c\simeq \frac{12\pi}{11 \left(1 - \frac{2N_f}{N_c}\right)
\ln\frac{\Lambda^2_C}{m^2_t}} > \frac{4\pi}{27} \quad \mbox{or} \quad
\Lambda_C \leq 10^4 GeV. \la{est}
\ee
Thereby with the help of CSR sum rules we have showed that there is a 
certain window to have relatively 
light composite Higgs bosons in the effective 2HQQM 
created by a more fundamental  QCD-like theory.

For a comparison we also check the consistency of the last sum rule \gl{last}
in the
one-channel top-condensate SM. One deals then with only one scalar,$H^0$, and 
a triplet of Goldstone bosons, $\pi^a$. Respectively, in the sum rules
\gl{constr}, \gl{last} one should retain only $Z^\pi, Z_{H^0}$. Hence they 
read:
\ba
Z^\pi &=& Z_{H^0}= \frac{N_c\Lambda^4}{16\pi^2\ln\frac{\Lambda^2}{m^2_t}};\no
  Z_{H^0} m^2_{H^0} &=& 
\frac14 \frac{N_c\Lambda_C^4 m^2_t }{\pi^2\ln\frac{\Lambda_C^2}{m^2_t}}
\,\not=\, \Delta_{SP} = 
\frac{9}{11}\frac{N_c\Lambda_C^4 m^2_t}{\pi^2 \ln\frac{\Lambda^2_C}{m^2_t}},
\la{onech}  
\ea 
where eqs.\gl{cor1} with $A \simeq A_{11}, B \simeq B_{11}$ from \gl{mA},
\gl{mB} are employed, the Nambu relation  $m^2_{H^0} = 4 m^2_t$ is used and
the correlators  are derived according to eqs.\gl{con1}, \gl{con2}. As well
the definition \gl{CSR} for $\Delta_{SP}$ and the expression \gl{est}
for $\alpha_s$ is inserted. Evidently there is a serious 
discrepancy between the left
and right sides of the last sum rule which is remarkably independent on
scales, colors and flavours. Thus it happens to be impossible to embed
the one-channel Top-mode SM into a QCD-like underlying theory.
\section{Conclusions}
\nn 1.\quad  We have found that in the framework of 2HQQM of type I two composite Higgs doublets can be created
dynamically as a consequence of DCSB in two channels. Lighter Higgs bosons appear as radial excited states in the 
language of the Potential Quark Model.\\

\nn 2.\quad  We have proved that CSR at high energies is realized in the 2HQQM of type I  in the Nambu-Jona-Lasinio
phase near tricritical point. Thereby these models in the NJL phase can be regarded as   
effective models originating from a QCD-like 
underlying theory with an electroweak compositeness 
scale of order $1 \div 10 TeV$ (it may be also a fundamental, string 
 scale \ci{banks} when, at least, additional five 
dimensions of inverse size 10 Mev exist). \\

\nn 3. \quad One can show that in other phases explored in two-channel models in \ci{aa,aay}, such as
the anomalous or special phase, the last CSR constraint cannot be fulfilled
for {\sl any} choice of parameters. It means that such phases cannot be realized in effective models
motivated by underlying QCD-like fundamental theory. But the question remains open about what kind of
underlying theory could induce such phases at lower energies.\\

\nn 4. \quad It turns out that the soft-momentum limit of the correlators
\gl{ans} is 
connected to the structural
constants of the effective chiral  lagrangian \ci{gas,bij} (in our case 
of the EW lagrangian
in  the large Higgs mass limit \ci{esp}):
\ba
\frac{Z_{\sigma}}{m^2_{\sigma}} + \frac{Z_{\sigma'}}{m^2_{\sigma'}}
 + C &=& 8 B^2_0 (2L_8 + H_2);\no
\frac{Z_{\pi'}}{m^2_{\pi'}}  + C &=&  8 B^2_0 (- 2L_8 + H_2).
\ea
One can think about how to apply these relations together with the empirical 
estimations on the corresponding structural coupling constant
for the determination of the size of 
the (techni-,top-)quark condensate
$<\bar q q> = - B_0 F_0^2$ ( with $F_0$ being an electroweak scale).\\

\nn 5. \quad  Since the first paper \ci{njl} on the compositeness 
due to strong 4-fermion forces it is known \ci{bard,marc,bar} that
the low-mass spectrum is provided by a fine-tuning. It is remarkable that the above pattern for heavy 
Higgs particles can be put in certain correspondence to the predictions
found with the help of the modified Veltman fine-tuning hypothesis \ci{apr}.\\

\nn 6. In this paper we have not considered the CP and electroneutrality 
violating scenario for
2HQQM in details although it may occur as an intermediate stage 
in the formation of baryon assymetry of the Universe \ci{lah,laine}.\\

\nn 7. \quad There exists another way to build multi-channel 
(nonlocal but separable) quark models
\ci{volk,aavy} and it  is of certain interest to develop  the similar 
CSR analysis in that approach.\\

\nn 8. \quad 
More realistic predictions for masses and
coupling constants  should
be based, of course, on the full SM action including vector bosons and on the RG improved calculations
of low-energy parameters both in Higgs and 
in fermion sectors which are in preparation.
\vspace{1cm}

\acknowledgments{
We express our gratitude to DFG for financial support which makes it possible
to continue our research on properties of composite bosons in
effective quark models. We are also  grateful to D. Espriu for useful 
discussion and invaluable remarks.}


\begin{thebibliography}{99}
\bi{ans}  A. A. Anselm, N. G. Uraltsev, V. A. Khoze,  
 {\it Sov. Phys. Usp.} {\bf 28} (1985) 113;\\
 A.A. Anselm, 
{\it Surveys High Energ.Phys.} {\bf 7} (1994) 107. 
\bi{lee} T.D.Lee, \prd{8}{1973}{1226}. 
\bi{gun} J.F.Gunion, H.E.Haber, G.Kane, S.Dawson,
{\it The Higgs Hunter's Guide}, Addison-Wesley 1990.
\bi{gw} S. L. Glashow, S. Weinberg, \prd{15}{1977}{1958}.
\bi{at} D. Atwood, L. Reina, A. Soni, \prd{55}{1997}{3156}.
\bi{krod} G. Kreyerhoff, R. Rodenberg,
\plb{226}{1989}{323};\\
J. Freund, G. Kreyerhoff, R. Rodenberg,
\plb{280}{1992}{267}.
\bi{sher} M. Sher, \prep{179}{1989}{273};\\
S. Nie, M. Sher, \hepph{9811234}.
\bi{lang} J. Erler, P. Langacker, \hepph{9809352}.
\bi{degr}G. Degrassi et al, \plb{418}{1998}{209};\\
G. D'Agostini, G. Degrassi, \hepph{9902226}.
\bi{njl} Y.Nambu, G.Jona-Lasinio, \pr{122}{1961}{345}.
\bibitem{bard}
 V. A. Miransky, M. Tanabashi, K. Yamawaki, \mpla{4}{1989}{1043}; \,
\plb{221}{1989}{177}.
\bi{marc} W. J. Marciano, \prl{62}{1989}{2793};
\,\prd{41}{1990}{219}.
\bi{bar}W.A.Bardeen, C.T.Hill, M.Lindner,\prd{41}{1990}{1647}.
\bi{ll} M. Lindner, D. L\"ust, \plb{272}{1991}{91};\\
 M. Lindner, \ijmpa{8}{1993}{2167}.
\bi{scm} A. A. Andrianov, V. A. Andrianov,  \hepph{9705364};\quad
 {\it Zapiski Nauch. Sem. POMI (Proc. Steklov Math. Inst., 
St. Petersburg, Russia)}, 
{\bf 245}, No.14 (1996) 5.
\bi{masha} M. Krawczyk, \hepph{9803484}; \, \hepph{9812536} and refs. therein.
\bibitem{aa} A.A.Andrianov and V.A.Andrianov,
\ijmpa{8}{1993}{1981};\\
{\it Theor. Math. Phys.} {\bf 93} (1992) 1126;\\ 
{\it Theor. Math. Phys.}{\bf 94} (1993) 3;\\ \hepph{9309297};\quad 
{\it Proc. School-Sem.
"Hadrons  and nuclei from QCD", Tsuruga/Vladivostok/Sapporo 1993}, WSPC (1994)
 341;\\
{\it Zap. Nauch. Sem. POMI (Proc. Steklov Math.Inst.,St. Petersburg, 
Russia )}, {\bf 209}, No.12 (1994) 3;\\ 
\npps{39B,C}{1995}{257}.
\bi{wk} K. G.  Wilson, J. B. Kogut,\prep{C12}{1974}{75}.
\bi{aary} A. A. Andrianov, V. A. Andrianov, V. L. Yudichev, R. Rodenberg,\\
\hepph{9709475} ;\\
\ijmpa{14}{1999}{323}. 
\bi{bark} T. L. Barklow {\it et al.}, {\it 1996DPF/DPB Summer Study, 
Snowmass, Colorado}, \hepph{9704217}.
\bi{tec}
T. Appelquist, J. Terning, L. C. R. Wijewardhana.  
\prl{79}{1997}{2767} and references 
therein.
\bi{top}  R. S. Chivukula, B. A. Dobrescu, H. Georgi, C. T. Hill. 
\hepph{9809470} and references therein.
\bi{aet} A. A. Andrianov, D. Espriu, R. Tarrach, \npb{533}{1998}{429}.
\bi{shif}M. A. Shifman, A. I. Vainstein, V. I. Zakharov,
\npb{147}{1979}{385, 448}.
\bibitem{cve}
G. Cvetic, \hepph{9702381} and references therein.
\bibitem{aay}
A.A.Andrianov, V.A.Andrianov, V.L.Yudichev, \, \hepph{9512256}, \quad
{\it Proc. X Int. Worksh. on HEP
and QFT (Zvenigorod, Russia, Sept. 1995)}, MSU Publ. (1996) 211;\,
\\\hepph{9610376},\quad
{\it Proc. XI Int. Worksh. on HEP
and QFT (St.Petersburg, Sept. 1996)}, MSU Publ. (1997) 160;\\
{\it Theor. Math. Phys.} {\bf 108} (1996) 1069.
\bi{cvek} G. Cvetic, C. S. Kim, S. S. Hwang, \prd{58}{1998}{116003}.
\bi{br}  G. C. Branco, \prd{22}{1980}{2901};\\
G.C. Branco, J.-M. Gerard, W. Grimus, \plb{136}{1984}{383}. 
\bi{osl} A. Skjold, P. Osland, \npb{453}{1995}{3};\\
C.A. Boe et al. \hepph{9811505}.
\bi{bern}  W. Bernreuter, \hepph{9808453} and refs. therein;\\
W. Bernreuter, A. Brandenburg, M. Flesch,\prd{56}{1997}{90};\,
\hepph{9812387}.
\bi{lah} A. B. Lahanas, V.C. Spanos, V. Zarikas, \hepph{9812535}.
 \bi{rod} R. Rodenberg, \ijmpa{11}{1996}{4779} \,
and references therein.
\bi{pdg} C.Caso {\it et al.} (PDG), {\it European Phys. J.} {\bf C3} (1998) 1.
\bi{nc} G. t'Hooft, \npb{72}{1974}{461}.
\bi{wit}E.Witten, \npb{160}{1979}{57}.
\bi{rein} L. J. Reinders, H. Rubinstein, S. Yazaki,
\prep{127}{1985}{1} and references therein.
\bi{jam} M. Jamin, M. M\"unz, \zpc{66}{1995}{633}.
\bi{rk} N. V. Krasnikov, R. Rodenberg, {\it Nuovo Cim.} {\bf A105} (1992) 1601.
\bi{banks} T. Banks, A. Nelson, M. Dine, \hepth{9903019}.
\bi{gas}J. Gasser, H. Leutwyler. \npb{250}{1985}{465}.
\bi{bij}J. Bijnens, E. de Rafael, H. Zheng, \zpc{62}{1994}{437};\\
J. Bijnens, \prep{265}{1996}{369}.
\bi{esp}A. Dobado, D. Espriu, M. J. Herrero, \plb{255}{1991}{405};\\
P. Ciafaloni , D. Espriu , \prd{56}{1997}{1752}.
\bi{apr} A.A.Andrianov, R. Rodenberg and N. V. Romanenko,
  {\it Nuovo Cim.} {\bf 108A} (1995) 577.
\bi{laine} M. Laine, K. Rummukainen, \hepph{9811369}.
\bibitem{volk}  M. K. Volkov, C. Weiss, \prd{56}{1997}{221};\\
M. K. Volkov, D. Ebert, M. Nagy, \hepph{9705334}.
\bi{aavy}A. A. Andrianov, V. A. Andrianov, M. K. Volkov, V. L. Yudichev,\\ 
   {\it JINR Rapid Commun.} {\bf 4 - 90} (1998) 45.
\end{thebibliography}
\end{document}